
\font\titlefont = cmr10 scaled \magstep2
\magnification=\magstep1
\vsize=22truecm
\voffset=1.75truecm
\hsize=15truecm
\hoffset=0.95truecm
\baselineskip=20pt

\settabs 18 \columns

\def\b{\bigskip}
\def\bb{\bigskip\bigskip}

\def\ce{\centerline}

\def\no{\noindent}

\rightline{AMES-HET 94-05}
\rightline{June 1994}
\bb

\ce{\titlefont{Phenomenology Of A Non-Standard Top Quark}}
\ce{\titlefont {Yukawa Coupling}}

\b

\ce{ X. Zhang, ~~S.K. Lee,~~ K. Whisnant ~and ~B.-L. Young}
\b
\ce{Department of Physics and Astronomy }
\ce{ Iowa State University,}
\ce{ Ames, Iowa
 50011}

\b
\bb
\ce{\bf ABSTRACT}

\no There are theoretical speculations that the top
quark may have different properties from that predicted by the standard
model.
We use an effective Lagrangian technique to model such a non-standard top
quark scenario. We parametrize
the CP violating interactions of the top quark with the bubble wall
 in terms of
 an effective top quark Yukawa coupling, then
  study its effects on the
electroweak baryogenesis.
 We also discuss the
phenomenology of such an effective Yukawa coupling
 in low and high energy regions.

\bb
\filbreak

\no {\bf {I. Introduction}}

The standard model (SM) is perfectly consistent with all evidence
gathered to date,
however, it raises as many questions as it answers. One important
 issue is that of mass generation. In the standard model, the
 fermion and gauge boson masses
 come from interactions
 which couple them to the symmetry breaking sector.
 Since the
top quark is heavier than all other observed fermions and
gauge bosons and has
  a mass of the order the electroweak
symmetry breaking scale, it couples to the symmetry breaking sector strongly.

The symmetry breaking sector of the standard model consists of
a complex fundamental scalar $\Phi$. However, there are the
 theoretical arguments of
 ``Triviality"[1] and ``Naturalness"[2] against such a scalar sector. So
one believes that the Higgs sector of the standard model is
an effective theory and expects that the new physics
 will
  manifest itself
in the effective interactions of the top quark. In this paper, we will study
the phenomenology of a non-standard top quark
Yukawa coupling in electroweak baryogenesis and in low
and high energy processes.

One can propose various models in which the top quark has
 a non-standard Yukawa coupling. For example,
consider the top quark condensate models[3]. In these models, the electroweak
symmetry breaking is realized by the top-antitop condensate
caused by a strong interaction involving top
quarks above the electroweak symmetry breaking scale.
 However, in the
original formulation of this idea,
a new strong interaction takes place at a very high energy and
 the effective theory in the TeV
 energy region
corresponds to the SM.
There are also
  models where the new physics which triggers the top quark condensation
 appears around
a few TeV. Since the new physics scale is of the order of
the Fermi scale, there would not be the hierarchy problem. And,
 more importantly,
one would expect many interesting phenomena
at the TeV energy scale. One of them
can be that the
effective Yukawa coupling, $\Gamma_t^{eff}$, differs from the SM.
Models with
 weak interactions
above the electroweak symmetry breaking scale
 can also produce an effective
Yukawa coupling $\Gamma_t^{eff}$
different from the SM. In the Appendix, we will give an example
of this latter approach
and derive $\Gamma_t^{eff}$ in a
class of the Left-Right (L-R) symmetric models.

 Given the fact that
  the underlying theory beyond
the SM is presently unknown
and so many disparate models exist in
the literature, it is worthwhile
to extract phenomenologically
the essence of the predictions in a ``fundamental"
symmetry breaking and mass generation theory by
means of an effective Lagrangian technique[4].
 In
this paper we will show
that if $\Gamma_t^{eff}$ violates CP symmetry it will
help to produce the observed baryon asymmetry at the weak scale.

 In the context of electroweak
baryogenesis, an effective Lagrangian approach has been taken in Ref.[5].
 Following Ref.[5],
we parametrize the new physics effects by a
set of higher
dimension operators ${\cal O}^i$:

$${
{\cal L}^{new} = \Sigma_i {c_i \over \Lambda^{d_i -4}} {\cal O}^i ~~~~, }
\eqno(1)$$

\no where $d_i$ are integers greater than 4. The
${\cal O}^i$, which have dimension $[ {\rm mass} ]^{d_i}$, are
invariant under the SM gauge symmetry and contain only the SM fields.
The parameters $c_i$, which determine the strength of the contribution
of ${\cal O}_i$, can in principal be calculated by
matching the effective theory with the underlying theory. However, they are
taken as free parameters here since we do not know the underlying theory.
$\Lambda$ is a mass parameter which is the scale at
 which the effective theory breaks down. Here
we shall take $\Lambda$ to be the order of a few TeV. Barring
the presence of anomalously large coefficient $c_i$, we expect that
${\cal L}^{new}$ will be dominated by low dimension operators.

The lowest dimension ${\cal O}^i$, which modifies
the Yukawa coupling and satisfies the above requirements, is given by

$${
{\cal O}^{t}= ~c_t ~e^{i \xi} ~{( \phi^2 - {v^2 \over 2} )\over \Lambda^2}
   ~ \Gamma_t ~{\overline{\Psi_L}} {\tilde \Phi}
t_R
{}~~~~~, }\eqno(2.a)$$

\no where ${\overline{\Psi_L}}\equiv
( {\overline{t_L}}, ~~{\overline{b_L}} ),$~
$\Gamma_t$ is the SM Yukawa coupling,
$\xi$ is a CP violation phase and $c_t$ is a real parameter.
Note that in our notation the $\Phi$ is an
$SU_L(2) \times U_Y(1)$ doublet scalar and
$\phi \equiv {({ \Phi^\dagger \Phi })}^{1\over 2}$.
 By definition, the operator in
Eq.(2.a)
does not renormalize the top quark mass $m_t = {\Gamma_t \over {\sqrt 2}}
{}~v$, but it
 modifies the top-Higgs interaction (see Eq.(7) below).
Combining (2.a) with the SM Yukawa coupling,
we have
 an effective Yukawa coupling for the top quark,

$${
\Gamma_t^{eff} = \Gamma_t ~ \{ 1 + c_t ~e^{i \xi}
                     {( \phi^2 - {v^2 \over 2} ) \over \Lambda^2}
                     \} ~~~~. }\eqno(2.b)$$
\b
In the following sections we will
examine in detail the phenomenology of the
effective Yukawa coupling $\Gamma_t^{eff}$
in electroweak baryogenesis and in relevant
  low and high energy processes.
The paper is organized as follows.
 In section II we estimate
 the coefficents in (2) by considering their effect on
 electroweak baryogenesis.
In section III, we calculate the electric dipole moments of fermions
induced by $\Gamma^{eff}_t$.
 In section IV, we consider the
phenomenological implication of the $\Gamma^{eff}_t$
 in a future Linear Collider.
 Section V includes a summary and discussion.

\b

\no {\bf {II. Baryogenesis With An Effective Top Quark Yukawa Coupling}}

The first order electroweak phase transition proceeds by the formation
of bubbles
of the true vacuum which grow and fill
the universe. The baryon asymmetry can be produced in
the bubble wall and
in front of it.
In the presence of a CP violating effective top quark Yukawa coupling
$\Gamma_t^{eff}$ (Eq.2.b),
the most efficient mechanism for baryogenesis is the charge transport
mechanism considered by Cohen, Kaplan and Nelson[6]. For this mechanism
to work, the
thickness of the bubble wall is required to be smaller than the typical
particle mean free path $\sim 4 / T$ for quarks[7],
where $T$ is the phase transition temperature. In general,
this requirement can be satisfied in electroweak models
 with $SU_L(2)$ singlet scalar fields[7]. One would also expect
that the
thin bubble wall
scenario will be applicable in the effective theory. As
argued in Ref.[8], an important effect of the new physics on
the effective potential is to shift the quartic coupling of the
Higgs field from $\lambda_T \sim m_H^2 / 2 v^2 $ to
$ \lambda_T \sim (m_H^2 / 2 v^2) - (4 v^2/ \Lambda^2)$.
Thus if $\Lambda$ is not too far from the Fermi scale, one can make
$\lambda_T$ small enough to have a thin wall without conflicting with the
experimental lower bound on the Higgs boson mass $m_H > 60$ GeV[9].

The asymmetry arising from the different
number of the left-handed and
right-handed top quarks which scatter off
the bubble wall
 serves as a force to bias the anomalous baryon
number violating reaction. However,
to compute the net baryon density produced, one needs to know the distribution
of the asymmetry in front of the bubble wall by solving numerically
 appropriate Boltzmann equations[10]. If the system is partially in
thermal equilibrium, however, the procedure could be simplified.

Approximately, the final ratio of baryon number density to
entropy can be expressed as

$${
{n_B \over s} \sim \kappa ~ \alpha_W^4 ~\delta_{CP} F ~~~~~, } \eqno(3)$$

\no where $F$
is a factor which
depends on the properties associated with the phase transition
and the $\kappa$ subsumes our ignorance about the sphaleron rate
 in the unbroken phase. Quantatively, $\kappa > 0.5$ as derived in
lattice simulation[11] and
$\kappa \sim 20$ as estimated in a different method given
in Ref.[12].
{}From Ref.[6], $F$ could be as large as of
the order of 0.1, which gives ${n_B \over s}
\sim 10^{-7}$ for
$\kappa \sim 1$ and a
maximal CP violation in the two-Higgs model. Here,
$\delta_{CP} \sim c_t \sin\xi v^2 / 2 \Lambda^2$, we would expect
that for $\Lambda = 1$ TeV,

$${
{n_B \over s} \sim \kappa ~c_t \sin\xi ~ 10^{-9} ~~~. } \eqno(4)$$

\no In order to explain the observed asymmetry ${n_B \over s} \sim
( 0.4 - 1.4 ) \times 10^{-10}$, it is required that,

$${
 \kappa~ c_t \sin\xi \geq 4 \times 10^{-2} ~~~~. } \eqno(5) $$

 We would like to point out that since
the electroweak baryogenesis calculations available so far are qualitative,
the result in (5) is only accurate to within a couple of orders of
magnitude. For instance, if the effect of the QCD sphaleron[13]
is taken into consideration,
 the asymmetry in (4) would be
suppressed by a factor $10^{-2}$[14][15].
 As a result, the CP violation required should be
larger than that shown in (5), namely,

$${
\kappa ~c_t \sin\xi \geq 4 ~~~~. }  \eqno(6)$$

\no{\bf {III. Electric Dipole Moments Of The Electron And Neutron}}

In the unitary gauge,
 $\Phi = {(v + H) \over{\sqrt 2}} \pmatrix{0 \cr 1\cr}$ with
$H$ being the physical Higgs particle, the effective Yukawa coupling
$\Gamma_t^{eff}$ generates a non-standard CP violating
 top-Higgs interaction,

$${
{\cal L}^{eff} \sim {m_t \over v} {\overline t}~ [
           (1 + ~({c_t \over 16})~  \cos\xi )
           + i  ~({c_t \over 16}) ~   \sin\xi ~ \gamma_5
               ]~ t ~~H ~~~~, } \eqno(7)$$
\no where we have fixed $\Lambda = 1$ TeV.

In the last section, we have put a lower bound on $\kappa c_t \sin\xi$
from baryogenesis. In this section, we will calculate the
contribution of ${\cal L}^{eff}$ to the electric dipole moment
of the fermion. A recent discussion of this kind
of calculation in a gauge theory is given in Ref.[16]. For
the electric dipole moment of the electron, $d_e$,
the dominant contribution is from the two-loop diagram[17]
in Fig.1. In the calculation, we first evaluate the top quark
triangle to get an effective
photon-photon-Higgs operator,
then we evaluate the one-loop diagram in Fig.2, and cut off the logarithmic
divergence by the top quark mass.
 We have

$${
{d_e \over e} \sim {m_e \over v} ~{( {1 \over 27 })}~ c_t \sin\xi~
              {\alpha_{em} \over \pi v} ~{1 \over 16 \pi^2 }
                   \ln{m_t^2 \over m_H^2 }
 ~~~~. } \eqno(8)$$

\no The value of the Higgs boson mass
is constrained in electroweak baryogenesis.
Its upper limit in the standard model is about
40 GeV[18].
In an effective theory, it can be relaxed to the
experimentally allowed region[8].
In the numerical calculation of
$d_e$, we take $m_H \sim 80$ GeV
and
$m_t \sim 174$ GeV[19]. Given the constraints on
the quantity $\kappa c_t \sin\xi$ as shown in (5)
and (6), we have the prediction on
the $d_e$ from the electroweak baryogenesis,
$${
{d_e \over e} \sim {6 \over \kappa} \times ( 10^{-28}
\sim 10^{-30} )~ cm ~~. }\eqno(9)$$

\no Clearly, for $\kappa$ in the range mentioned above, the
electric dipole moment of the electron is just
below the current experimental upper bound. This
 opens the
possibility of detecting $d_e$ in the near future.

The estimate of the electric dipole moment of
the neutron $d_n$ contains large uncertainties. So we will not go into
this calculation in detail. As an estimate, we use
 the nonrelativistic quark model to
calculate the electric dipole moments of the up and down quarks,
$d_u ~{\rm and}~ d_d$,
 using
 a diagram similar to that in Fig.1. We have
$d_n = {4\over 3} d_d -{1\over 3} d_u  \sim {m_d \over m_e } d_e
\sim {1 \over \kappa}\times ( 2 \times 10^{-26} \sim 8 \times 10^{-29} )
 ~e. cm $. Thus the electric dipole
moment of the neutron is also within the experimental limit.

Before concluding this section, we would like to examine the
contribution of ${\cal L}^{eff}$ to the QCD $\theta$, which is given by
$${
{\cal L}^{\theta} = {\overline\theta} ~{g_s^2 \over 32 \pi^2 }~
  G^{\mu\nu} {\tilde G}_{\mu\nu} ~~~~, }\eqno(10)$$

\no where ${\overline \theta} = \theta + {\rm arg ~det ~M}$, with
$M$ being the mass matrix of the quark fields. Strong interactions,
through the ${\overline \theta}$ dependence, lead to
an estimate for $d_n$. The
experimental bound on $d_n$ implies that[20]
$${
{\overline \theta } < 10^{-9}~~~. } \eqno(11)$$

\no The relevant diagram we should evaluate is the
  top quark self energy correction
  shown in Fig.3. In our calculation,
 the dimensional regularization is used and the infinity
associated with the phase of the top quark mass
( $ {\rm arg }~ m_{t}$ )
 induced by loop effects
 can be absorbed in the
counterterms appearing in the effective Lagrangian.
The coefficient of this counterterm is a new independent coupling in
the effective theory and can not be computed. So in what followes,
we shall focus on the "log-enhanced" term and
compute the leading correction by setting the counterterm to zero
for the renormalization scale to be $\Lambda (= 1 {\rm TeV})$,
 the cut-off of the theory.
 Thus we have

$${ \eqalign{
{\rm arg }~ m_{t} & \sim   c_t \sin\xi ~{v^2 \over \Lambda^2 }~
           {2 \over 16 \pi^2 }~ {m_t^2 \over v^2 } \ln{\Lambda^2 \over m_t^2}
           \cr
       & \sim {3 \over \kappa } \times ( 10^{-4} \sim 10^{-6} ) ~. \cr }
 } \eqno(12)$$

\no Clearly, if no cancellation occurs
between the original $\theta$ and
the induced $\theta
= {\rm arg } ~ m_{t}$,
 an axion[20] must be included in the effective lagrangian.

\b

\no {\bf IV. Phenomenology Of The Top Quark Yukawa Coupling At Linear Collider}

There is a direct way to determine the Yukawa coupling
from top and Higgs production at a high energy
collider[21]. In this
section, we consider the possibility of
testing the top quark Yukawa coupling
 at future $e^+ e^-$ colliders.
For a light Higgs boson,
$m_H < 120$ GeV, as required by electroweak baryogenesis,
the dominant process for top pair plus Higgs boson production is
 the bremsstrahlung process in Fig.4.
 For the standard model coupling
of the Higgs boson to top quark, the cross section has been calculated
in Ref.[22]. They concluded that for an integrated
luminosity of $\int {\cal L} = 20 ~ {\rm fb}^{-1}$, some 100 events can
 be expected at Higgs
masses of order 60 GeV, falling to 20 events at 120 GeV.
Even though the production rate is low, as argued in Ref.[22], the
signature of the process $e^+ e^- \rightarrow t {\overline t} H
\rightarrow W^+ W^- b {\overline b} b {\overline b}$ is spectacular,
so that it is not impossible to isolate the events experimentally.

Given the effective lagrangian (7) for the Top-Higgs coupling, the Dalitz plot
density
for $e^+ e^- \rightarrow t {\overline t} H$ can be written as

$${
{ d \sigma ( e^+ e^- \rightarrow t {\overline t} H ) \over
{ dx_1 ~dx_2 }} = \{ { d \sigma \over {dx_1 ~dx_2}} \}_{SM}~
(1+ 2 \delta \cos\xi
           +\delta^2 ) -  \{ { d \sigma \over {dx_1 ~dx_2} } \}_{BSM}~
              ( \delta \sin\xi )^2 , }
\eqno(13)$$
\no where $\delta = {c_t / 16}$.
The
$\{ { d \sigma \over {dx_1 ~ dx_2 }} \}_{SM}$ is given in Eq.(7)
of Ref.[22].
And our calculation gives

$${
\eqalign{
\{ { d \sigma \over {dx_1~dx_2} } \}_{BSM}= & 4 N_c {\sigma_0 \over {4 \pi}}
                    {g_{t t H}^2 \over {4 \pi }}
              \{ G_+ ~{f \over x_{12} } [ h - x + {x^2 \over x_{12}}
                     ( 1-f ) ] \cr
              & + G_- ~{3 f \over x_{12}} [ x - h + {x^2 \over x_{12}}f ] \}
            ,  \cr}
}\eqno(14)$$

\no where $g_{t t H} = {m_t \over v}, ~~\sigma_0 = {4 \pi \alpha^2 \over
         { 3 s}}, ~~h = {m_H^2 \over s}, ~~f = {m_t^2 \over s},~~
        x_{12}=(1- x_1 ) ( 1 - x_2 ), ~~ x_1 = 2 E_t/ {\sqrt s}, ~~
            x_2 = 2 E_{ \overline t}/ {\sqrt s}, ~~x=2 E_{H}/ {\sqrt s},~~
N_c = 3$ is the number of colors, and

$${
G_{\pm} = Q_e^2 Q_t^2 + { 2 Q_e Q_t {\hat v}_e {\hat v}_t \over
{1 - M_Z^2 / s} } + { ({\hat v}_e^2 + {\hat a}_e^2) ( {\hat v}_t^2
             \pm {\hat a}_t^2 ) \over { ( 1 - M_Z^2  / s )^2 } }
{}~~~~, } \eqno(15.a)$$

\no with
$${
{\hat v}_t = { 1 - 4 Q_t x_W \over {4 {\sqrt{x_W(1-x_W)}}}} , ~~
       {\hat a}_t = {1 \over { 4 {\sqrt{x_W(1-x_W)}}}} ~~,
} \eqno(15.b)$$
\no and $x_W = \sin^2\theta_W \simeq 0.23$.

To illustrate the effects of the non-standard top quark Yukawa coupling
in Eq.(2.b)
on the production of the top and Higgs,
in Fig.5 we plot in the $\delta - \xi$ parameter space contours of
${\sigma / \sigma_{SM} }$, the ratio of the integrated
 cross section to the SM cross section,
 for
$m_t = 174$ GeV and
$m_H = 80$ GeV.
The $\sigma / \sigma_{SM} = 0.60 ~{\rm and} ~ 1.40$
 contours represent the parameters for which the
integrated cross section is 3 standard deviations (considering
statistical uncertainty only) from the SM value, assuming an
 integrated luminosity of 20
$fb^{-1}$. Regions above these contours should therefore
 be readily distinguishable
from the SM. The dotted line gives the
lower bound on $\delta$ from
electroweak baryogenesis from
 Eq.(6) with $\kappa = 0.5$, which is the
 most restrictive limit using the indicated ranges of
parameters in Sec. II. At the other extreme, using Eq.(5) with
$\kappa = 20$ gives no substantial bounds on $\delta$
and
$\xi$.

\bb
\b
\no {\bf V. Conclusion And Discussions}

In this paper we have examined the impact of a non-standard top quark
on electroweak baryogenesis.
 We parametrize the CP violating interaction of the top
quark with the bubble wall in terms of an effective
Yukawa coupling[23] (Eq.(2.b)).
 We found that to explain the observed baryon number
asymmetry, the electric dipole moments of the electron and neutron must be
very close to the present experimental limits. We have explored
 the possibility of
testing the effective Yukawa coupling in the future linear collider and
concluded that its effect on the cross section of the top and Higgs production
is sizable (see Fig.5).
 However, we would like to point out that the
 electroweak baryogenesis calculations
available so far are qualitative and the quantitative results
 obtained are probably
only accurate to within a couple of orders of magnitude.
For instance, the uncertainty of $\kappa$ in the sphaleron
 rate generates an uncertainty in the predictions
on $d_e ~{\rm and}~ d_n$.
Thus future experimental inputs will test the present knowledge about
electroweak baryogenesis.

\b
\bb
This work is supported in part by the Office of High Energy
 and Nuclear Physics
of the U.S. Department of Energy (Grant No. DE-FG02-94ER40817).

\bb
\vfill\eject

\ce{ \bf Appendix}

In this Appendix we illustrate how to generate
the operator ${\cal O}_t$ in a weakly interacting Higgs
model. Let us examine
the effective Lagrangian of the
  left-right
symmetric (L-R) models at an energy
 below the $SU_R(2)$ breaking scale. Specifically we consider a class of
L-R models described in Ref.[24], where the Higgs sector consists
of a complex bidoublet $\phi ( 2, 2, 0)$ and
one set of L-R symmetric lepton-number-carrying triplets $\Delta_L (3, 1, 2)
\bigoplus \Delta_R (1, 3, 2)$. In the fermion sector, there are
L-R symmetric doublets of three generations of the quarks and leptons.

After the $SU_R(2)$ is broken at the scale
$V_R$, the right-handed neutrinos receive Majorana masses
$\sim V_R$ and decouple at
low energy. However, they
 do have effects on low energy physics,
such as making the left-handed
neutrinos light {\it via} a see-saw mechanism. In terms of an effective
Lagrangian below $V_R$, the neutrino majorana mass terms can be parametrized
by an $SU_L(2) \times U_Y(1)$ gauge invariant higher dimension operators.
Now we will show that an higher dimension operator ${\cal O}^t$ can
 be generated by heavy Higgs scalars in this model.

The bidoublet $\phi$ will be split into two
$SU_L(2) \times U_Y(1)$ doublets below the scale
$V_R$, which we denote by
$\phi_1$
and $\phi_2$. The top quark Yukawa sector can be rewritten in terms of
Higgs fields $\phi_1 ~{\rm and}~ \phi_2$ by

$${
{\cal L}^{top} = h_1 {\overline \Psi}_L \phi_1 t_R ~+~ h_2 {\overline \Psi}_L
                   {\tilde \phi_2} t_R ~~, } \eqno(16)$$
\no where $h_1 ~{\rm and}~ h_2$ are Yukawa couplings in the L-R symmetric
Lagrangian. The vacuum expectation values of $\phi_1 ~{\rm and}~ \phi_2$
are related to that of the bidoublet $\phi$. For a general
CP violating Higgs potential, one has,

$${
< \phi > = e^{i \alpha} \pmatrix{ \kappa & 0 \cr
                                  0 & \kappa^\prime \cr}
          ~~~, } \eqno(17)$$

\no where $\alpha$ is a CP phase. So
$< \phi_1 > = e^{i \alpha} \kappa , ~~ < \phi_2 > = e^{i \alpha}
\kappa^\prime$.
Thus the mass of the top quark is given by
$${
m_t = h_1 \kappa e^{i \alpha} ~+~ h_2 \kappa^\prime e^{-i \alpha}
{}~~~. } \eqno(18)$$

\no The detailed Higgs potential involving $\phi_1 ~{\rm and}~ \phi_2$
can be obtained from that of $\phi , ~\Delta_R ~{\rm and}~ \Delta_L$.
Here we give a few terms which are relevant to our discussion,
$${
V( \phi_1, \phi_2) = |\lambda_1|e^{-i \delta} V_R^2 \phi_1 \phi_2
                 + \lambda_2 ( \phi_1 \phi_2 )^2 + ... ~~, } \eqno(19)$$

\no where $\lambda_1 ~{\rm and}~ \lambda_2$ are two coupling constants,
and $\delta$ is a CP phase.

After $SU_L(2) \times U_Y(1)$ is broken, there are two
``extra" neutral scalars as well
as a charged scalar
in addition to the neutral scalar corresponding to the Higgs
boson of the SM.
The experimental constraints[25] on the flavor-changing neutral currents
require these ``extra" scalars be heavier than a few TeV. However,
this in general does not need any fine tuning,
for $ V_R > 1$ TeV, in accordence with
 the extended survival hypothesis[26]: Higgs bosons acquire the maximum mass
compatible with the pattern of the symmetry breaking.
 So the effective theory below
$V_R$ contains only
a doublet scalar.
 At this point, we would like to
 point out that Haber and Nir[27] have made
a detailed study on the low energy structure of the multi-Higgs models
with a cut-off $\Lambda$.
  They concluded
 that if performing minimally the required fine-tuning
in order to set the electroweak scale $v$, the low energy scalar spectrum
is identical to that of the SM, up to corrections of order $v^2
\over \Lambda^2$. Here,
$\Lambda \sim V_R$.

To examine the effects of heavy scalars on the low energy
physics in term of $SU_L(2)
\times U_Y(1)$ invariant effective operators, we follow the procedure
in Ref.[27], and derive the effective Lagrangian by using
 a ``rotated" basis
$\{ \Phi , \Phi^\prime \}$, so that
$< \Phi > = v = {\sqrt{\kappa^2 + {\kappa^\prime}^2 }} ~{\rm and}~
< \Phi^\prime > = 0$.
{}From eq.(17), one has
$${
\pmatrix{\Phi \cr \Phi^\prime \cr}
= e^{- i \alpha} \pmatrix{\cos\zeta & \sin\zeta \cr
                        - \sin\zeta & \cos\zeta \cr}
       \pmatrix{ \phi_1 \cr \phi_2 \cr }
{}~~~, } \eqno(20)$$

\no where $\tan\zeta = \kappa^\prime / \kappa$.
In this basis, $\Phi$ serves as the SM doublet and
the $\Phi^\prime$ is just an massive scalar with mass $\sim V_R$.
Integrating out the heavy field $\Phi^\prime$
 will generate many higher dimension operators[28]. The contribution of a
Feynman diagram in Fig.6
  will give an operator
similar to ${\cal O}^t$. In this model,
$ m_{\Phi^\prime}^2 \sim V_R^2 \sim \Lambda^2$,
 the parameter $c_t$ and
CP phase $\xi$ in (2),
 as functions of the couplings $h_1 ,~ h_2, ~\lambda_2$,
the mixing angle $\zeta$ and the CP phases $\alpha , ~\delta$,
are calculable. However, a detailed examination of the
 whole structure
for the effective Lagrangian
 and the calculation of the various coeffients in the
effective Lagrangian of the L-R models are beyond the scope of this paper.

\bb
\bb
\vfill\eject

\bb
\ce{\bf Figure Captions}

\b
\item{[Fig.1]} Dominant contribution to $d_e$, the electric
 dipole moment of the electron.

\item{[Fig.2]} Same as Fig.1 with the fermion loop replaced
 by an effective photon-photon-Higgs coupling.

\item{[Fig.3]} Higgs boson contribution to the top
quark self energy.

\item{[Fig.4]} Dominant contribution to $e^+ e^- \rightarrow
         H t {\overline t}$. Another graph where H couples to
${\overline t}$ is also present.

\item{[Fig.5]} Contours of $\sigma / \sigma_{SM}$ in
the $\delta - \xi$ plane for
$m_t = 174$ GeV and
$m_H = 80$ GeV.
The region with $1 \leq \xi  / \pi \leq 2$ is a mirror image
of the region shown.
 The dotted line indicates the lower bounds on $\delta$ from
electroweak baryogenesis from Eq.(5) with
$\kappa = 0.5$.

\item{[Fig.6]} A possible diagram for the generation of ${\cal O}^t$
in eq.(2) in left-right symmetric models.

\bb
\b

\vfill\eject

\ce{\bf References}
\b
\item{[1]}For reviews, see, D. Callaway, Phys. Rept. 167, 241 (1988);
H. Neuberger, in the proceeding of the XXVI International
Conference on HIGH ENERGY PHYSYCS, Vol. II, P1360 (1992) ed. J. Sanford.

\item{[2]}G. 't Hooft, in Recent Developments in
Gauge Theories, ed. G. t'Hooft (Plenum Press, New York, 1980).

\item{[3]}For reviews, see, W. Bardeen,
 in {\it{Proceedings of the XXVI International
     Conferences on HIGH ENERGY PHYSICS}}, Gallas, Texas, U.S.A. 1992,
ed. J.R. Sanford;
M. Linder, Inter. J. Mod. Phys A, Vol. 8, 2167 (1993).

\item{[4]}For recent reviews on electroweak effective Lagrangian, see,
for example, F. Feruglio, Int. Jour. of Mod. Phys. A8, 4937 (1993);
J. Wudka, University of California at Riverside Preprint, UCRHEP-T121 (1994).

\item{[5]}Xinmin Zhang and Bing-Lin Young, Phys. Rev. D49, 563 (1994).

\item{[6]}A. Cohen, D. Kaplan and A. Nelson, Nucl. Phys. B373, 453 (1992).

\item{[7]}For a review, see, A. Cohen, D. Kaplan and A. Nelson,
          Ann. Rev. Nucl. Part. Sci. j43, 27 (1993);

\item{[8]}
X. Zhang, Phys. Rev. D47, 3065 (1993).

\item{[9]}see, for example, G. Coignet, Plenary talk at the
XVI International Symposium on Lepton-Photon Interaction,
Cornell University, Ithaca, New York, 10-15 August 1993.

\item{[10]}M. Joyce, T. Prokopec and N. Turok, PUPT-91-1437 (1993),
hep-ph/9410352.

\item{[11]}J. Ambjorn, T. Askgaad, H. Porter and
M. Shaposhnikov, Phys. Lett. 244B, 479 (1990).

\item{[12]}P. Arnold and L. McLerran, Phys. Rev. D36, 581 (1987).

\item{[13]}L. McLerran, E. Mottola and M. Shaposhnikov, Phys. Rev. D43,
           2027 (1991); R.N. Mohapatra and
 X. Zhang, Phys. Rev. D45, 2699 (1992).

\item{[14]}G. Giudice and Shaposhnikov, Phys. Lett. B326, 118 (1994).

\item{[15]}The observation that
in the massless limit
the force of biasing the sphaleron reaction vanishes
can be found in [14] and also in footnote 4 of a earlier work by
R.N. Mohapatra and X. Zhang, Phys. Rev. D46, 5331 (1992).

\item{[16]}Darwin Chang, Fermilab-Conf-90/265-T, NUHEP-TH-90/38, Dec. 20
(1990).

\item{[17]}S. Barr and A. Zee, Phys. Rev. Lett. 65, 21 (1990).

\item{[18]}M. Shaposhnokov, Nucl. Phys. B287, 757 (1987); B299, 797 (1988);
M. Dine, P. Huet and R. Singleton Jr, Nucl. Phys. B375, 625 (1992);
M. Dine, P. Huet, R. Leigh, A. Linde and D. Linde, Phys.
Rev. D46, 550 (1992).

\item{[19]}CDF Collaboration, F. Abe et al, FERMILAB-PUB-94/097-E.

\item{[20]}For a review on strong CP problem, see, R.D. Peccei,
in {\it CP violation} ed. by C. Jarskog (World Scientific Publishing
Co. Singapore, 1989).

\item{[21]}Testing the Yukawa coupling at Hadron Collider has been mentioned
in, C.-P. Yuan, Michigan State University Preprint, MSUHEP-93/10, July 1993.

\item{[22]}A. Djouadi, J. Kalinowski and P.M. Zerwas,
Mod. Phys. Lett. A7, 1765 (1992); Z. Phys. C54, 255 (1992).

\item{[23]}The use of an effective Yukawa coupling to test CP violation in
top quark production at high energy colliders has been considered,
 {\it e.g.}, in D. Chang, W.-Y. Keung, and I. Phillips, Nucl. Phys. B408,
  286 (1993); Phys. Rev. D48, 3225 (1993); W. Bernreuther and A. Brandenburg,
              Phys. Lett. B314, 104 (1993).

\item{[24]}R.N. Mohapatra and G. Senjanovi\'c, Phys. ReV. Lett. 44, 912 (1980);
       Phys. Rev. D23, 165 (1981).

\item{[25]}R.N. Mohapatra, G. Senjanovi\'c, and
M. Tran, Phys. Rev. D28, 546 (1983); G. Ecker, W. Grimus, and H. Neufeld,
Phys. Lett. B120, 365 (1983).

\item{[26]}H. Georgi and D.V. Nanopoulos,
Phys. Lett. B82, 95 (1979); F. del Aguila and L.E. Ib\'a\~nez,
 Nucl. Phys. B177, 60 (1981); R.N. Mohapatra and
 G. Senjanovi\'c, Phys. Rev. D27, 1601 (1983).

\item{[27]}H. Haber and Y. Nir, Nucl. Phys. B335, 363 (1990).

\item{[28]}The operator
${\phi^2\over \Lambda^2}{\tilde W}W$,
where $W$ is
$SU_L(2)$ field strength,
 considered
for electroweak baryogenesis
in M. Dine, P. Huet, R. Singleton Jr and L. Susskind,
 Phys. Lett. B257, 351 (1991),
(see also Ref.[5]) does not appear in the effective Lagrangian here. Such a
 operator needs heavy fermion loop. In this model, no heavy
 fermions,
except the right-handed neutrinos, are present.

\bye

\bye